\documentclass[10pt,letterpaper,dvipdfmx]{article}
\usepackage[top=0.85in,left=2.75in,footskip=0.75in]{geometry}

\usepackage{amsmath,amssymb,amsthm}

\usepackage{changepage}

\usepackage[utf8x]{inputenc}

\usepackage{textcomp,marvosym}

\usepackage{cite}

\usepackage{nameref,hyperref}

\usepackage[right]{lineno}

\usepackage{microtype}
\DisableLigatures[f]{encoding = *, family = * }

\usepackage[table]{xcolor}

\usepackage{array}

\usepackage[colorinlistoftodos]{todonotes}

\newcolumntype{+}{!{\vrule width 2pt}}

\newlength\savedwidth

\usepackage{ulem}
\usepackage{color}

\newcommand{\one}[1]{\mathbf{1}_{#1}}

\newcommand{\argmin}{\mathop{\rm arg~min}\limits}

\newcommand{\Com}[1]{#1}

\newcommand{\DEL}[1]{}
\newcommand{\ADD}[1]{#1}
\newcommand{\REP}[2]{\DEL{#1}\ADD{#2}}

\newtheorem{lem}[]{Lemma}

\raggedright
\usepackage[aboveskip=1pt,labelfont=bf,labelsep=period,justification=raggedright,singlelinecheck=off]{caption}

\makeatletter
\renewcommand{\@biblabel}[1]{\quad#1.}
\makeatother

\date{}

\usepackage{lastpage,fancyhdr,graphicx}
\usepackage{epstopdf}
\fancyheadoffset[L]{2.25in}
\fancyfootoffset[L]{2.25in}
\begin{document}
\vspace*{0.2in}

\begin{flushleft}
{\Large
\textbf\newline{Fast and exact search for the partition with minimal information loss} 
}
\newline
\\
Shohei Hidaka\textsuperscript{1},
Masafumi Oizumi\textsuperscript{2,3},
\\
\bigskip
\textbf{1}  Japan Advanced Institute of Science and Technology, Nomi-shi, Ishikawa, Japan
\\
\textbf{2} Araya Inc., Minato-ku, Tokyo, Japan
\\
\textbf{3} RIKEN Brain Science Institute, Wako-shi, Saitama, Japan
\\
\bigskip

%
%





*shhidaka@jaist.ac.jp, oizumi@araya.org

\end{flushleft}
\section*{Abstract}
In analysis of multi-component complex systems, such as neural systems, identifying groups of units that share similar functionality will aid understanding of the underlying structures of the system. To find such a grouping, it is useful to evaluate to what extent the units of the system are separable. Separability or inseparability can be evaluated by quantifying how much information would be lost if the system were partitioned into subsystems, and the interactions between the subsystems were hypothetically removed. A system of two independent subsystems are completely separable without any loss of information while a system of strongly interacted subsystems cannot be separated without a large loss of information. Among all the possible partitions of a system, the partition that minimizes the loss of information, called the Minimum Information Partition (MIP), can be considered as the optimal partition for characterizing the underlying structures of the system. Although the MIP would reveal novel characteristics of the neural system, an exhaustive search for the MIP is numerically intractable due to the combinatorial explosion of possible partitions. Here, we propose a computationally efficient search to precisely identify the MIP among all possible partitions by exploiting the {\it submodularity} of the measure of information loss\ADD{, when the measure of information loss is submodular}. Submodularity is a mathematical property of set functions which is analogous to convexity in continuous functions. Mutual information is one such submodular information loss function, and is a natural choice for measuring the degree of statistical dependence between paired sets of random variables. By using mutual information as a loss function, we show that the search for MIP can be performed in a practical order of computational time for a reasonably large system ($N = 100 \sim 1000$). We also demonstrate that MIP search allows for the detection of underlying global structures in a network of nonlinear oscillators.

\section{Introduction}
The brain can be envisaged as a multi-component dynamical system, in which each of individual components interact with each other. One of the goals of system neuroscience is to identify a group of neural units (neurons, brain area, and so on) that share similar functionality \cite{Humphries2011,Lopes-dos-Santos2013,Carrillo2016,Romano2017}. 

Approaches to identify such functional groups can be classified as ``external’’ or ``internal’’. In the external approach, responses to external stimuli are measured under the assumption that a group of neurons share similar functionality if their responses are similar. A vast majority of studies in neuroscience have indeed used the external approach, by associating the neural function with an external input to identify groups of neurons or brain areas with similar functionality \cite{Hubel1962}.
 
On the other hand, the internal approach measures internal interactions between neural units under the assumption that neurons with similar functionality are connected with each other. The attempts to measure internal interactions have rapidly grown following recent advancements in simultaneous recording techniques \cite{Harris2003, Schneidman2006, Stevenson2011}. It is undoubtedly important to elucidate how neurons or brain areas interact with each other in order to understand various brain computations. 
\DEL{In particular, to understand awareness or consciousness, the internal approach is preferable to the external approach because consciousness cannot be essentially associated with any external signals. For example, dreaming is a phenomenon which clearly tells us that consciousness occurs regardless of any external signals.} 

\DEL{Consistent with this, the Integrated Information Theory (IIT) of consciousness is constructed to focus on the internal interactions in a neural system. 
IIT states that the prerequisite for consciousness is integration of information realized by the internal interactions of neurons in the brain. IIT proposes to quantify the degree of information integration by an information theoretic measure, ``integrated information'' and hypothesizes that integrated information should be related to the level of consciousness.} 

\begin{figure}[t]
  \includegraphics[width=0.95\linewidth]{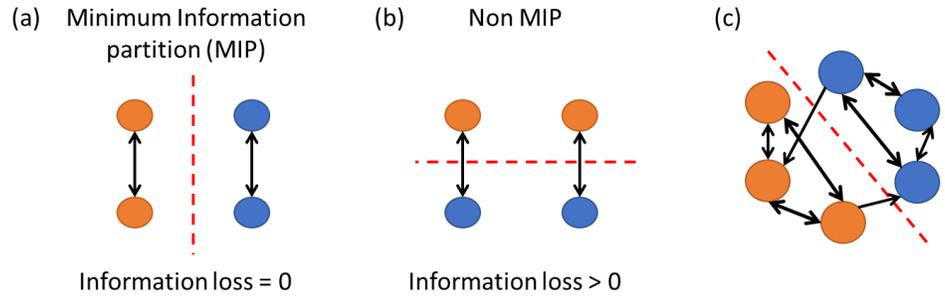} 
  \caption{(a) Minimum information partition (MIP). (b) Another possible partition, which differs from the MIP. (c)  MIP in a general network where there is no clear-cut partition. }
  \label{fig:MIP}
\end{figure}

\DEL{Conceptually, integrated information quantifies the degree of interaction between parts or, equivalently, the amount of information loss caused by splitting a system into parts.} 
\DEL{By using a measure of information loss such as integrated information, we can find functional groups of neural units using the criterion of ``minimal information loss''.} 

\ADD{In this study, we consider the problem of finding functional groups of neural units using the criterion of ``minimal information loss''. Here, ``information loss'' refers to the amount of information loss caused by splitting a system into parts, which can be quantified by the mutual information between groups.} For example, consider the system consisting of 4 neurons shown in Fig. \ref{fig:MIP}(a). The two neurons on each of the left and right sides are connected with each other but do not connect with those on the opposite side. The natural inclination is to partition the system into left (orange) and right (blue) subsystems, as shown in Fig. \ref{fig:MIP}(a). This critical partition can be identified by searching for the partition where information loss is minimal, \ADD{i.e., mutual information between the two parts is minimal}. 
\DEL{In IIT, the partition with minimal information loss is called the ``Minimum Information Partition (MIP)''.}
In fact, if a system is partitioned with MIP as in the example system, \REP{integrated information }{ information loss (mutual information between the subsystems)} is 0 because \REP{there is no information loss }{there are no connections between the left (orange) and the right (blue) subsystems}. If the system is partitioned in a different way than MIP, as shown in Fig. \ref{fig:MIP}(b), \REP{integrated information}{information loss} is non-zero because there are connections between the top (orange) and the bottom (blue) subsystems. This is not the optimal grouping of the system from the viewpoint of information loss. 

\DEL{With the MIP approach, we can quantify ``global'' integration in a whole system rather than ``local'' integration in subsystems. The system in Fig. \ref{fig:MIP} is considered to be a system in which information is locally integrated within each left or right subsystem but is not globally integrated between the left and right subsystems. Integrated information with the MIP shown in Fig. \ref{fig:MIP}(a) quantifies global integration, which is zero and integrated information with the partition shown in Fig. \ref{fig:MIP}(b) quantifies local integration, which is non-zero. IIT considers that a system's global integration determines consciousness as a whole system. Aside from its relevance to consciousness, MIP could also be used to identify functional groups of neural units from the viewpoint of global information loss. }

\ADD{The concept of the partition with minimal information loss originated from Integrated Information Theory (IIT) \cite{Tononi2004,Balduzzi2008,Oizumi2016} and the partition with minimal information loss is called ``Minimum Information Partition (MIP)''. In IIT, information loss is quantified by integrated information \cite{Tononi2004,Balduzzi2008,Oizumi2016}, which is different from the mutual information we use in this study. Although the measure of information loss is different, we use the same technical term ``MIP'' in this study as well because the underlying concept is the same. }

Although the theoretical idea of MIP is attractive to \REP{research into consciousness and other fields of}{ the fields of neuroscience} as well as to network science in general, \ADD{it has been difficult to apply it to the analysis of large systems.} In a general case in which there is no obvious clear-cut partition (Fig. \ref{fig:MIP}(c)), an exhaustive search for the MIP would take an exceptionally large computational time which increases exponentially with the number of units. This computational difficulty has hindered the use of MIP-based characterization of a system. 

In this study, we show that the computational cost of searching for the MIP can be reduced to the polynomial order of the system size \REP{if a measure of integrated information is a submodular function}{by exploiting the submodularity of mutual information}. We utilize one of the submodular optimization, the Queyranne's algorithm \cite{Queyranne1998}, and show that the exponentially large computational time is drastically reduced to $O(N^{3})$, where $N$ is the number of units, when we only consider bi-partitions. We also extend the framework of the Queyranne's algorithm to general $k$-partition and show that the computational cost is reduced to $O(N^{3(k-1)})$. The algorithm proposed in this study is an exact search for the MIP, unlike previous studies which found only the approximate MIP \cite{Tegmark2016,Toker2016}. This algorithm makes it feasible to find MIP-based functional groups in real neural data such as multi-unit recordings, EEG, ECoG, etc.., which typically consist of $\sim 100$ channels. 

The paper is organized as follows. In the Section \ref{sec-problem}, we formulate the search for the MIP, and show that mutual information is one of the submodular functions, and that we can treat it as a measure of information loss for a bi-partition. In Section \ref{sec-application}, we report on numerical case studies which demonstrate the computational time of this MIP search for analysis of a system-wise correlation and also demonstrate its use for analysis of a nonlinear system. In Section \ref{sec-discussion}, we discuss the potential use of the submodular search for other measures which are not exactly submodular.

\section{Methods \label{sec-problem}}
\subsection{Submodular function}
For a ground set $\mathcal{X} = \{ x_{1}, \ldots, x_{N}\}$ of $N$ elements and any pair of subsets $X, Y \in 2^{\mathcal{X}}$, if a set function $f: 2^{\mathcal{X}} \mapsto \mathbb{R}$ holds the inequality
\begin{equation}
\label{eq-Submodular1}
 f( X ) + f( Y ) \ge f( X \cup Y ) + f( X \cap Y ),
\end{equation} 
we call it {\em submodular} (See \cite{Iwata2008} for a review of submodularity). Equivalently, for $X \subseteq Y \subseteq Z$ and $z \in Z \setminus Y$, a submodular set function $f$ holds 
\begin{equation}
\label{eq-Submodular2}
 f( X \cup z ) - f( X ) \ge f( Y \cup z ) - f( Y ).
\end{equation}
If $-f( X )$ is submodular, we call it {\em supermodular}. 

Submodularity in discrete functions can be considered as an analogue of convexity in continuous functions. 
\ADD{
Intuitively, Eq. (\ref{eq-Submodular1}) means that in some sense the sum of two components scores higher than the whole. Eq. (\ref{eq-Submodular2}) means when something new is added to a smaller set, it has a larger increase in the function than adding it to a larger set. Also, the reader will be able to have the intuitive idea behind these inequalities by considering the special case, when the equality holds for {\it modular} function, that is both submodular and supermodular. For example, the cardinality of a set $f(X) = |X|$ is modular, and holds equality for both Eqs. (\ref{eq-Submodular1}) and (\ref{eq-Submodular2}).
}

\ADD{It is easy to find the equivalence between the inequality Eq. (\ref{eq-Submodular1}) and Eq. (\ref{eq-Submodular2}): Apply Eq. (\ref{eq-Submodular1}) to $X' = X \cup z$ and $Y$ such that $X \subseteq Y \subseteq Z$ and $z \in Z \setminus Y$, and we have (\ref{eq-Submodular2}). For converse, assume $X \subseteq Y \subseteq Z$ and $z \in Z \setminus Y$, and apply Eq. (\ref{eq-Submodular2}) to 
a series of paired sets, $X_{0}:= X$ and $Y_{0}:=Y$, and $X_{i}:= X_{i-1} \cup z_{i}$ and $Y_{i}:=Y_{i-1} \cup z_{i}$ for every $0 < i \le |Z \setminus Y|$ and $z_{1}, z_{2}, \ldots \in Z \setminus Y$. Then, we have Eq. (\ref{eq-Submodular1}) by summing up these series of inequalities.
}

It has been shown that the minimization of submodular functions can be solved in polynomial order of computational time, circumventing the combinatorial explosion. In this study, we utilize submodular optimization to find the partition with minimal information loss (Minimum Information Partition (MIP)). 

\subsection{Minimum Information Partition (MIP)}
We analyze a system with $N \in \mathbb{N}$ distinct components. Assume that each of the $N$ components is a random variable, and denote the random variable of the $i^{\text{th}}$ components by $x_{i}$ for $1 \le i \le N$. Denote the set of indices $\mathcal{N}:=\{1, 2, \ldots, N\}$ and the set of the $N$ variables by $V = \{ x_{1}, \ldots, x_{N}\}$. For the sake of simplicity, we consider bipartition of the whole system $V$ for the explanatory purpose. We will deal with a general $k$-partition in Section \ref{sec:recursive}. $V$ is divided into two parts $M$ and $\overline{M}$ where $M$ is a non-empty subset of the whole system $V$, $M \subset V$ and $\overline{M}$ is the complement of $M$, i.e., $\overline{M}= V \setminus M$. Note that bipartition $(V \setminus M, M)$ of a fixed set $V$ is uniquely determined by specifying only one part, $M$, because the other part is determined as the complement of $M$. Minimum Information Partition (MIP), $M_{\rm MIP}$, is defined as the subset that minimizes the information loss caused by partition, indicated by a non-empty subset $M \subset V$,
\begin{equation}
M_{\rm MIP}(V) := \argmin_{M \subset V, M \neq \emptyset} f(M), \label{eq:MIP_def}
\end{equation}
where $f(M)$ is the information loss caused by a bipartition specified by the subset $M$. More precisely, ``MIP'' defined in Eq. \ref{eq:MIP_def} should be called ``Minimum Information Bipartition (MIB)'' because only bi-partition is taken into consideration. However, as we will show in Section \ref{sec:recursive}, the proposed method is not restricted only to a bi-partition and can be extended to a general $k$-partition. To simplify terminology, we only use the term ``MIP'' through out the paper even when only bi-partition is considered. \DEL{IIT proposes several measures for information loss, which it terms ``integrated information''.} 
\DEL{In terms of integrated information, MIP is defined as the partition where integrated information between the parts is minimized. }

The number of possible bi-partitions for the system size $N$ is $2^{N-1}-1$, which grows exponentially as a function of the system size $N$. Thus, for even a modestly large number $N$ of variables ($N \sim 40$), exhaustively searching all bi-partitions is computationally intractable.

\subsection{Information loss function}
In this study, we use the mutual information between the two parts $M$ and $\overline{M}$ as an information loss function,
\begin{align}
f(M) &:= I\left(M;\overline{M} \right), \\
&= H(M) + H\left(\overline{M}\right) - H\left(M,\overline{M}\right), \label{eq:MI}
\end{align}
where $H(X)$ is the Shannon entropy \cite{CoverThomas1991,Shannon1948} of a random variable $X$, 
\[
 H( X ) := -\sum_{ x \in X} P( x )\log P( x ).
\]
As we will show in the next section, the mutual information is a submodular function. The mutual information (Eq. \ref{eq:MI}) is expressed as the KL-divergence between $P(V)$ and the partitioned probability distribution $Q(V)=P(M)P(\overline{M})$ where the two parts $M$ and $\overline{M}$ are forced to be independent,
\begin{equation}
I(M;\overline{M}) = D_{KL} \left( P(X)||P(M)P\left(\overline{M} \right) \right). \label{eq:KL_MI}
\end{equation}
The Kullback-Leibler divergence measures the difference between the probability distributions and can be interpreted as the information loss when $Q(V)$ is used to approximate $P(V)$ \cite{Burnham2003}. Thus, the mutual information between $M$ and $\overline{M}$ (Eq. \ref{eq:KL_MI}) can be interpreted as information loss when the probability distribution $P(V)$ is approximated with $Q(V)$ under the assumption that $M$ and $\overline{M}$ are independent \cite{Oizumi2016}. 

\DEL{In the first version of IIT (IIT 1.0), the mutual information was used as a measure of integrated information.}
\DEL{The difference between the mutual information considered in this study and that in IIT 1.0 is that the maximum entropy distribution is used as a probability distribution of one part $M$ or $\overline{M}$. Although different measures that take account of the dynamical aspects of a system are proposed in the later versions of IIT,} 
\DEL{we do not consider them in this study (see Discussion).} 
\DEL{Also, note that in IIT, the measures of integrated information are based on a perturbational approach, which attempts to quantify actual causation by perturbing a system into all possible states.} 
\DEL{The perturbational approach requires the full knowledge of the physical mechanisms of a system, i.e., how the system behaves in response to all possible perturbations. Aiming at an empirical application of our method, we only consider the measure (Eq. \ref{eq:KL_MI}) based on an observational distribution that can be estimated from empirical data.}

\subsection{Submodularity of the loss functions}
We will show that the mutual information (Eq. \ref{eq:MI}) is submodular. To do so, we use the submodularity of entropy. The entropy $H(X)$ is submodular because for \Com{$X \subset Y \subset Z$} and $z \in Z \setminus Y$, 
\begin{align*}
 H( X \cup \{z\} ) - H( Y \cup \{z\} ) &= - H( Y \setminus X \mid X \cup \{z\} ) \\
&\ge - H( Y \setminus X \mid X ) \\
&= - H( Y ) + H( X ),
\end{align*}
which satisfies the condition of submodularity (Eq. \ref{eq-Submodular2}).

By straightforward calculation, we can find that the following identity holds for the loss function $f(M)=I(M;\overline{M})$.
\begin{multline}
f(A \cup B)+f(A \cap B) -f(A) -f(B) = H(A \cup B)+H(A \cap B) -H(A) -H(B) \\
+H(\overline{A} \cup \overline{B})+H(\overline{A} \cap \overline{B}) -H(\overline{A}) -H(\overline{B}).
\end{multline}
Thus, from the submodularity of the entropy, the following inequality holds,
\begin{equation}
f(A \cup B)+f(A \cap B) -f(A) -f(B) \leq 0,
\end{equation}
which shows that $f(M)=I(M;\overline{M})$ is submodular. 

\subsection{MIP search algorithm}
\DEL{2.5 Recursive search for $k$-partition} \label{sec:recursive}

A submodular system \Com{$(V, f)$} is said to be {\em symmetric} if \Com{$f( M ) = f( V \setminus M)$} for any subset \Com{$M \subseteq V$}. It is easy to see that the mutual information is a symmetric submodular function from Eq. \ref{eq:MI}. When a submodular function is symmetric, the minimization of submodular function can be solved more efficiently. Applying Queyranne's algorithm \cite{Queyranne1998} \DEL{(See also Appendix in Section \ref{sec-Queyranne})}, we can precisely identify the bi-partition with the minimum information loss in computational time $O(N^{3})$. \ADD{See also Supporting Information 1 for more detail of the Queyranne's algorithm.}

\Com{\DEL{As introduced above, Queyranne's algorithm can find the minimum information bi-partition.} We can extend the \ADD{Queyranne's} algorithm \ADD{for bi-partition} to the exact search for a general $k$-partition with minimal information loss although it is more computationally costly. In what follows, we specifically explain $3$-partition case for simplicity, but the argument is applicable to any $k$-partition in a form of mathematical induction. } \ADD{See also Supporting Information 2 for more detail of the extension of the bi-partition algorithm.}

\DEL{Denote by the set of all $k$-partitions $(M_{0}, M_{1}, \ldots, M_{k-1})$ of $V$ by $P_{k,V}$. For the set $P_{k,V}$ and $2 \le k \le |V|$, define the function $g: P_{k,V} \mapsto \mathbb{R}$ by
$g((M_{0}, M_{1}, \ldots, M_{k})) := \sum_{i=1}^{k} H( M_{i} ) - H( V ),$
and it gives a natural extension of the information loss function to the $k$-partition. For the special case $k=2$, the reader would fine this corresponds with the function $g( (M_{0}, M_{1} ) ) = I( M_{0}; V \setminus M_{0} )$, mainly discussed in this study.
This extension of the mutual information is known as total correlation or multi-information.} 

\Com{
}

\DEL{2.6 Sequential MIP search}

\DEL{As shown in the previous section, the exact search for MIP in the case of a general $k$-partition is computationally costly. Moreover, for the moment, it is unclear which criterion we should employ to choose the $k$-partition solution over $(k-1)$-partition one. In practical application, thus, one may want to apply the bi-partition procedure sequentially, which is much faster than the exact minimum $k$-partition, and easy to decide the partition size -- partitioning until no more set found to make a further partition.} \DEL{Thus, we implement {\it sequential MIP search}, in which each of subsets of the MIP found in a given set of variables is further partitioned at each recursive step until it satisfies certain condition (e.g., a given set size limit, a number of maximum number of steps, and so on). In Study 2 (Section \ref{sec-seqMIP}), we demonstrate a sequential MIP search until it reaches the step with no further possible partition. Note that the sequential MIP search introduced here does not coincide with the exact minimum $k$-partition in general and is considered to be a crude approximation. However, it would be useful for a practical application because of the low computational costs.}

\section{Numerical Studies \label{sec-application}}
To demonstrate this search for the bi-partition with the minimal loss of information, we report here several case studies with artificial datasets. Throughout these case studies, we assume that the data is distributed normally. Under this assumption, we obtain the simple closed form 
\begin{equation}
f( M ) = I\left( M; \overline{M} \right) = \log_{2} |\Sigma_{M}| + \log_{2} |\Sigma_{\overline{M}}| - \log_{2} |\Sigma_X|,
no\end{equation}
where $\Sigma_X$ is the covariance matrix of the data, $\Sigma_{M},\Sigma_{\overline{M}}$ is the covariance matrix of the variables in the subsets $M$ and $\overline{M}$, and $|\Sigma|$ denotes the determinant of the matrix $\Sigma$. The computation of $|\Sigma_X|$ can be omitted because $|\Sigma_X|$ is constant across every step in the search and has no effect on the minimization of $I\left( M; \overline{M} \right)$.

\subsection{Study 1: Computational time}
In the first case study, we compare the practical computational time of the submodular search with that of the exhaustive search. We artificially generated a dataset consisting of 10,000 points normally distributed over $N$ dimensional space for $N = 2, 3, \ldots, 400$. Each dimension is treated as an element in the set. The exhaustive search is performed up to $N = 16$, but could not run in a reasonable time for the dataset with $N=17$ or larger due to limitations of the computational resource. Up to $N=16$, we confirmed that the submodular search found the correct MIPs indicated by the exhaustive search.

\begin{figure}[htbp]
  \begin{center}
   \includegraphics[width=.9\linewidth]{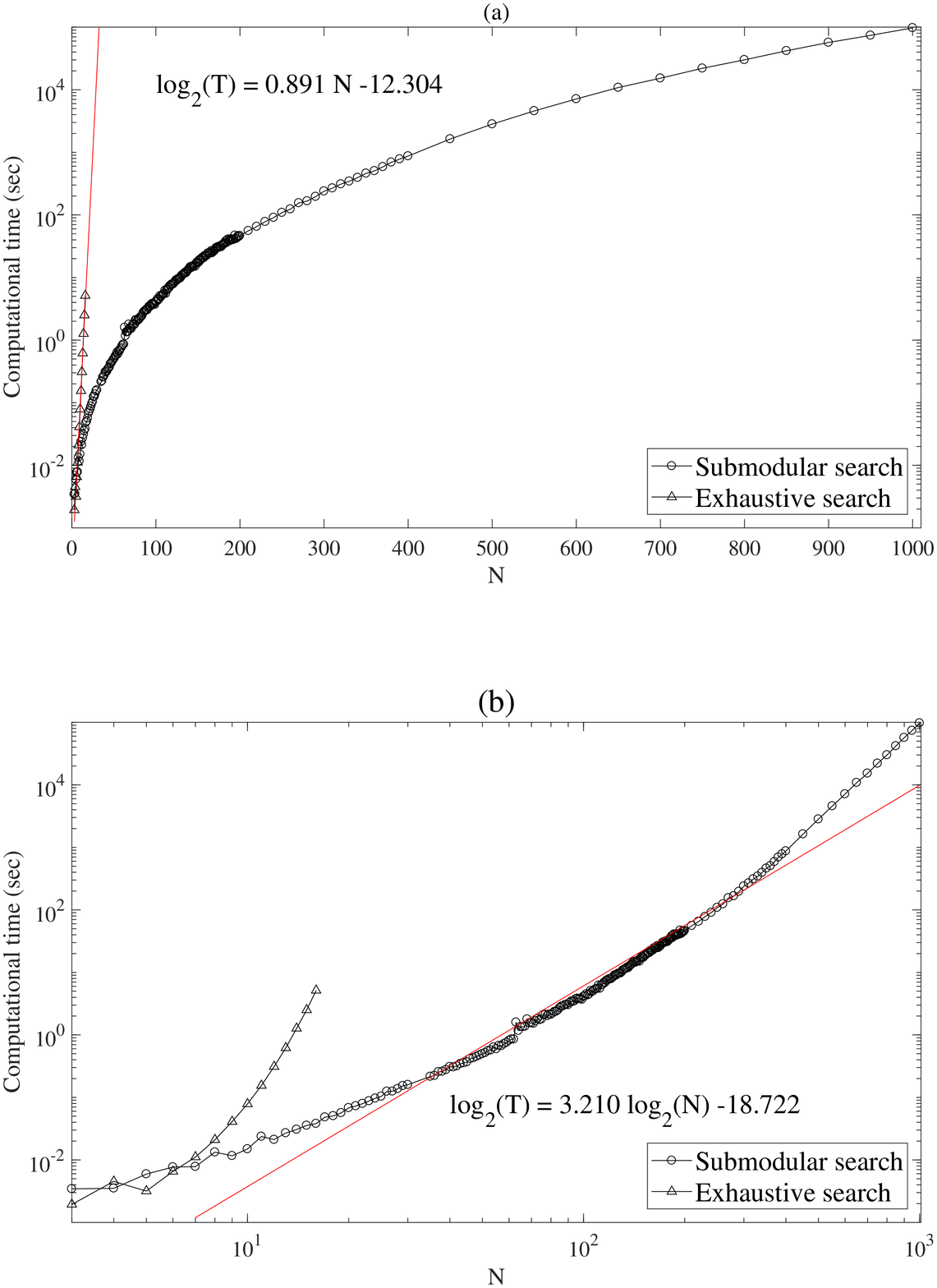}
  \end{center}
  \caption{(a) The semi-logarithm plot and (b) \REP{double-logarithm}{log-log plot} of the computation time for the two searches.}
  \label{fig-CompTime1}
\end{figure}

Figure \ref{fig-CompTime1} (a) shows the semi-logarithm plot of the computation time of the two searches. The empirical computation time of the exhaustive search was closely along the line, \Com{$\log_{2} T = 0.891 N - 12.304$}. This indicated that the exhaustive search took an exponentially large computational time $\approx O(2^N)$, which fits with the number of possible bi-partitions. Figure \ref{fig-CompTime1} (b) shows exactly the same results as the double-logarithm plot. In this plot, the computational time of the Queyranne's search was closely along the line $\log_{2}T = 3.210 \log_{2} N - 18.722$, which indicated that the Queyranne's search took cubic time $\approx O(N^3)$, as expected from the theory. With $N=1000$, the Queyranne's search takes 9738 seconds of running time. The computational advantage of the Queyranne's search over the exhaustive search is obviously substantial. For example, even with a modest number of elements, say $N=40$, the computational time of the exhaustive search is estimated to be $1.07 \times 10^7{\rm{sec}} \approx {\rm{123 days}}$ while that of the Queyranne's search is only $1 {\rm{sec}}$.

\subsection{Study 2: Toy example \label{sec-seqMIP}}
\begin{figure}[t]
  \includegraphics[width=\linewidth]
{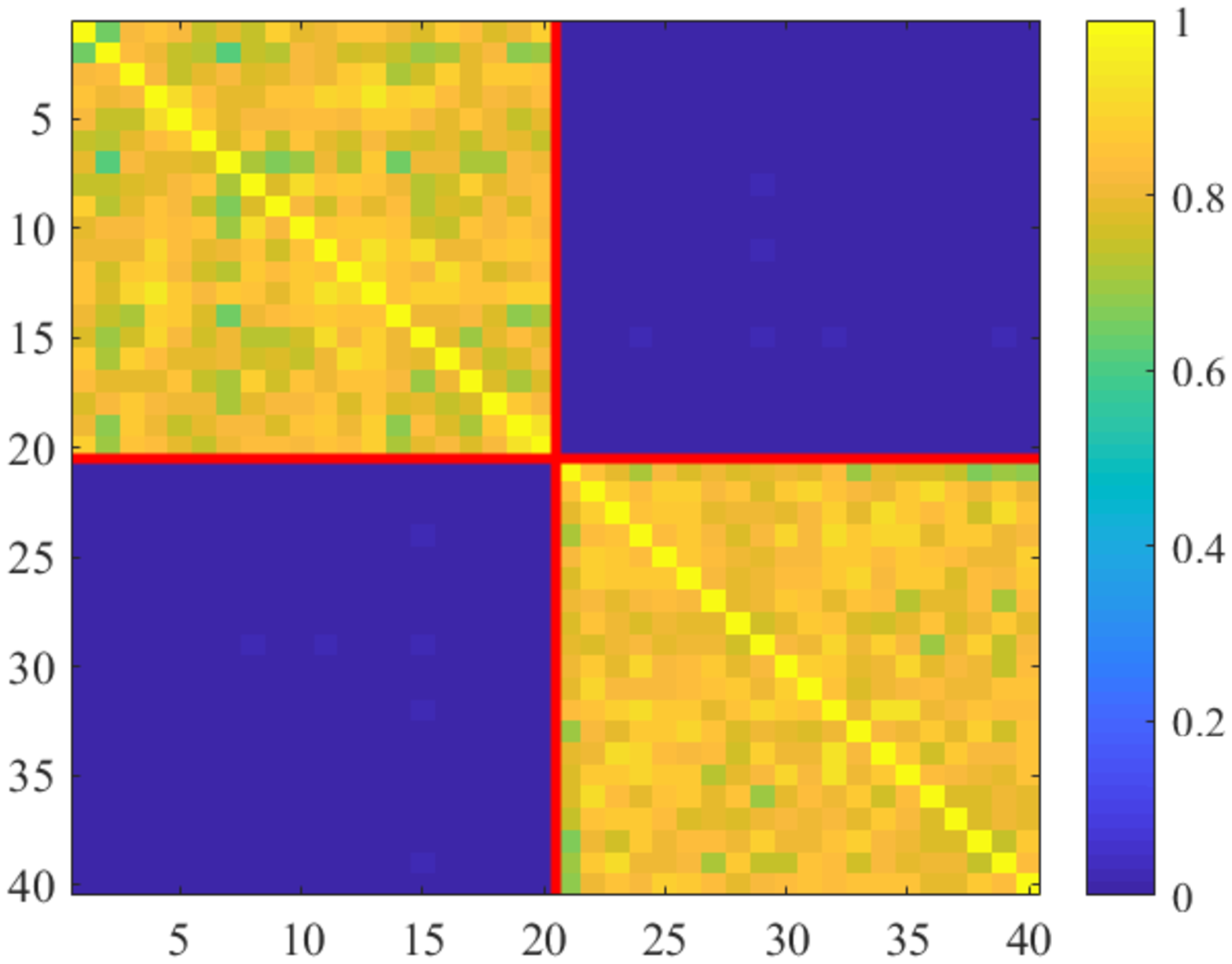}
\caption{\REP{Correlation matrix which reflects the toy problem with four correlated pairs in a system with eight components. The colored line shows partitioning at each step.}{A correlation matrix which reflects two subsets of variables $\{1, 2, \ldots, 20\}$ and $\{21, 22, \ldots, 40\}$.}}
\label{fig-toyproblem}
\end{figure}

As a demonstration of \DEL{sequential} MIP search \DEL{(Section 2.6)}, we consider a set of \REP{8}{40} random variables with the correlation matrix shown in Figure \ref{fig-toyproblem}. There are \REP{four pairs}{two subsets}, variables \REP{1 and 2}{1, 2, $\ldots$, 20}, \ADD{and} \REP{3 and 4, 5 and 6, and 7 and 8, with positive correlations, while any other pairs of variables show nearly zero correlation}{21, 22, $\ldots$, 40, within each subset with positive correlations, while any other pairs of variables across the two subsets shows nearly zero correlation}. 
\ADD{This simulated dataset is supposed to capture the situation visualized in Figure \ref{fig:MIP}(a) and (b). If the MIP search is successful, it would find the bipartition shown in Figure (a), in which each partitioned subset is either $\{1, 2\}$ or $\{3, 4\}$.}

The simulated correlation matrix is constructed as follows: \REP{For each pair of correlated variables (such as the Variable 1 and 2), a set of 1000 samples $(X, Y)$ is drawn using the independent bi-variate normal distribution with the mean $(0, 0)$ and the variance $(1, 0.5)$, and the coordinate $(X, Y)$ is rotated by $\pi/4$ to $(X', Y')$. \Com{This rotation was applied so that it lets the generated uncorrelated normally-distributed samples be correlated.}
}{
We first generate two matrices $X_{1}, X_{2} \in \mathbb{R}^{1000 \times 20}$ in which each of their elements  is a normally distributed random value. 
For $i=1,2$, let $U_{i}S_{i}V_{i}^{T} = X_{i}$ be the singular value decomposition of the matrix $X_{i}$, and 
construct another matrix $Y_{i} := U_{i}S_{i}( \lambda \one{20,20} + (1-\lambda)\epsilon_{20,20} )$, where
$\lambda = 0.1$, 
$\one{k,k}$ is a $k \times k$ matrix with all element being 1, and $\epsilon_{k,k}$ is a $k \times k$ matrix with each element being a normally distributed random value. 
}
The \REP{eight}{forty} dimensional dataset analyzed is constructed by concatenating the \REP{four pairs of variables}{$Y = (Y_{1}, Y_{2}) \in \mathbb{R}^{1000 \times 40}$}, each of which is constructed in this way.
By applying the \DEL{first}  MIP search, the system is partitioned into the pair of Variables \REP{1 and 2}{1, 2, $\ldots$, 20}, and the rest\ADD{, as expected (the red line in Figure \ref{fig-toyproblem} indicates the found MIP for the dataset)}. 

\DEL{As the pair of variables has no further meaningful partition, the other set of six variables is further partitioned. At this second step, the remaining six variables are partitioned into the pair of variables 3 and 4 and the rest. By recursively applying this procedure to the subset with a larger number of variables, we obtain the four subsets (1, 2), (3, 4), (5, 6), and (7, 8). As expected from Figure \ref{fig-toyproblem}, the sequential partitioning successfully re-discovers the underlying four groups of elements from the dataset.} 

\subsection{Study 3: Nonlinear dynamical systems}
In Study 3, we demonstrate how MIP changes depending on underlying network structures. For this purpose, we chose a nonlinear dynamical system in which multiple nonlinear components are chained on a line. Specifically, we construct a series of variants of the Coupled Map Lattice (CML) \cite{Kaneko1992}. Kaneko \cite{Kaneko1992} analyzed the CML in which each component is a logistic map and interacts with the one or two other nearest components on a line, and showed the emergence of multiple types of dynamics in the CML. In this model, each component is treated as a nonlinear oscillator, and the degree of interaction between other oscillators can be manipulated parametrically. By manipulating the degree of interaction, we can continuously change the global structure of the CMLs from one coherent chain to two separable chains. We apply the MIP search for the CMLs with different interaction parameters, and test whether the MIP captures this underlying global structure of the network.

\begin{figure}
\includegraphics[width=\textwidth]{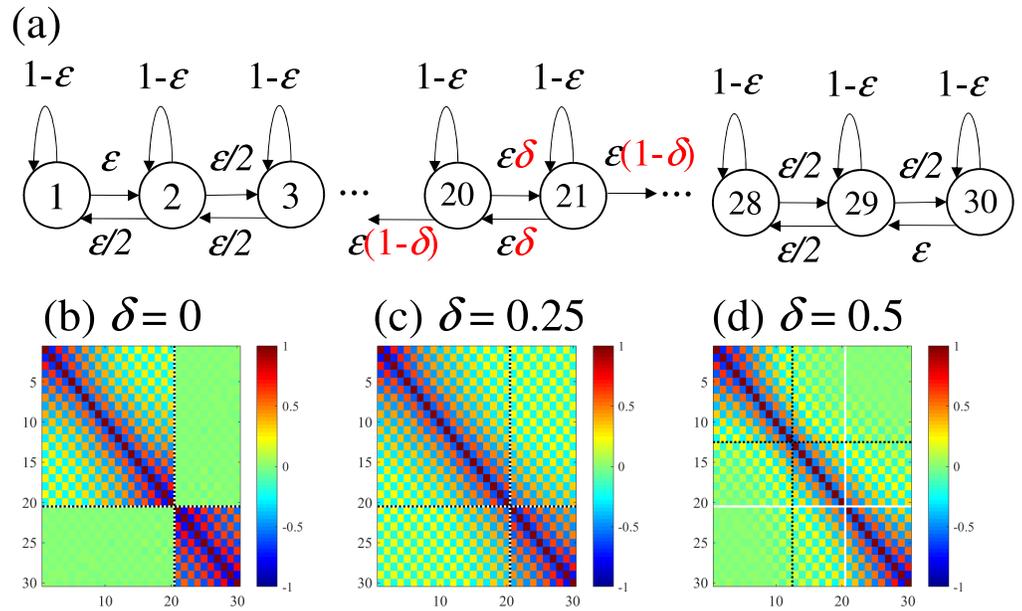}
\caption{\label{fig-nonlinearcorr} (a) The Coupled Map Lattice model with connection parameter $\delta$ indicating the connectivity between variables 20 and 21. The correlation matrices with (b) $\delta = 0$ (disconnected), (c) $\delta = 0.25$ (half-connected), and (d) $\delta = 0.5$ (fully connected). The white crossing lines show the expected partition as the ground truth at which the parameter $\delta$ is manipulated, and the black dotted crossing lines show the MIP typically found for each particular parameter.}
\end{figure}

Specifically, the CML is defined as follows. Let us write the logistic map with a parameter $a$ by $f_{a}( x ) := 1 - a x^2$. Let $x_{i,t} \in [0, 1]$ be a real number indicating the $i^{\text{th}}$ variable at $t^{\text{th}}$ time step for $i =1, 2, \ldots, N, t = 0, 1, \ldots, T$. For each $i = 1, \ldots, N$, the initial state of the variable $x_{i, 0}$ is set to a random number drawn from the uniform distribution on $[0, 1]$. For $t > 0$, we set the variables with the lateral connection parameter $\epsilon \in [0, 1]$ by 
\begin{eqnarray}
\nonumber
x_{1,t} &=& ( 1 - \epsilon ) f_{a}( x_{i, t-1} ) + \epsilon f_{a}( x_{i+1, t-1} ),
\\
\nonumber
 x_{N,t} &=& ( 1 - \epsilon ) f_{a}( x_{i, t-1} ) + \epsilon  f_{a}( x_{i-1, t-1} ),
\\
\nonumber
 x_{i,t} &=& ( 1 - \epsilon ) f_{a}( x_{i, t-1} ) + \frac{ \epsilon }{ 2 } \left( f_{a}( x_{i-1, t-1} ) + f_{a}( x_{i+1, t-1} ) \right) \text{ for } 1 < i < N.
\end{eqnarray}
According to the previous study \cite{Kaneko1992}, the defect turbulence pattern in the spatio-temporal evolution in $(x)_{i,t}$ is observed with the parameter $a = 1.8950$ and $\epsilon = 0.1$. In this study, we additionally introduce the ``connection'' parameter between the variables $i = 20, 21$ among $N=30$ variables (Figure \ref{fig-nonlinearcorr}(a)). Namely, with the connection parameter $\delta$, we redefine variables 19, 20, 21 and 22 by 
\begin{eqnarray}
x_{19,t} &:=& ( 1 - \epsilon ) f_{a}( x_{19, t-1} ) + \frac{ \epsilon }{2} f_{a}( x_{18, t-1} ) + ( 1 - \delta )\epsilon f_{a}( x_{20, t-1} )
\nonumber
\\
x_{20,t} &:=& ( 1 - \epsilon ) f_{a}( x_{20, t-1} ) + \frac{ \epsilon }{2} f_{a}( x_{19, t-1} ) + \delta \epsilon f_{a}( x_{21, t-1} )
\nonumber
\\
 x_{21,t} &:=& ( 1 - \epsilon ) f_{a}( x_{21, t-1} ) + \frac{ \epsilon }{2} f_{a}( x_{22, t-1} ) + \delta \epsilon f_{a}( x_{20, t-1} ) 
\nonumber
\\
 x_{22,t} &:=& ( 1 - \epsilon ) f_{a}( x_{22, t-1} ) + \frac{ \epsilon }{2} f_{a}( x_{23, t-1} ) + (1 - \delta) \epsilon f_{a}( x_{21, t-1} ) .
\nonumber
\end{eqnarray}
With the connection parameter $\delta = 1/2$, this model is identical to the original CML above, and with $\delta = 0$, it is equivalent to the two independent CMLs of $(x_{1}, \ldots, x_{20})$ and $(x_{21}, \ldots, x_{30})$, as it has no interaction between variable 20 and 21.

Given a sufficiently small connection parameter $0 \le \delta  < 1/2$, we expect that the MIP would separate the system into the subsets $\{1,2,\dots,20\}$ and $\{21,22,\dots,30\}$because the degree of the interaction between units 20 and 21 is the smallest. On the other hand, if the system is fully connected, which happens when $\delta = 1/2$, we expect that the  MIP would separate the system into the subsets $\{1,2,\dots,15\}$ and $\{16,17,\dots,30\}$ (in the middle of 30 units), due to the symmetry of connectivity on the line. The correlation matrices for different connection parameters $\delta$ from 0 to 1/2 are shown in Figure \ref{fig-nonlinearcorr}. In \Com{each matrix of \ref{fig-nonlinearcorr} (b), (c) and (d)}, the crossed \Com{white} lines show the expected separation between variables 20 and 21 \Com{at which the parameter $\delta$ is manipulated}. We found the block-wise correlation patterns in the matrix with $\delta = 0$, as expected (a typically found partition is depicted in the black dotted line in Figure \ref{fig-nonlinearcorr} (b), (c) and (d)); similar but less clear patterns with $\delta = 0.25$; and no clear block-wise patterns with $\delta = 1/2$.

\begin{figure}
  \includegraphics[width=\textwidth]{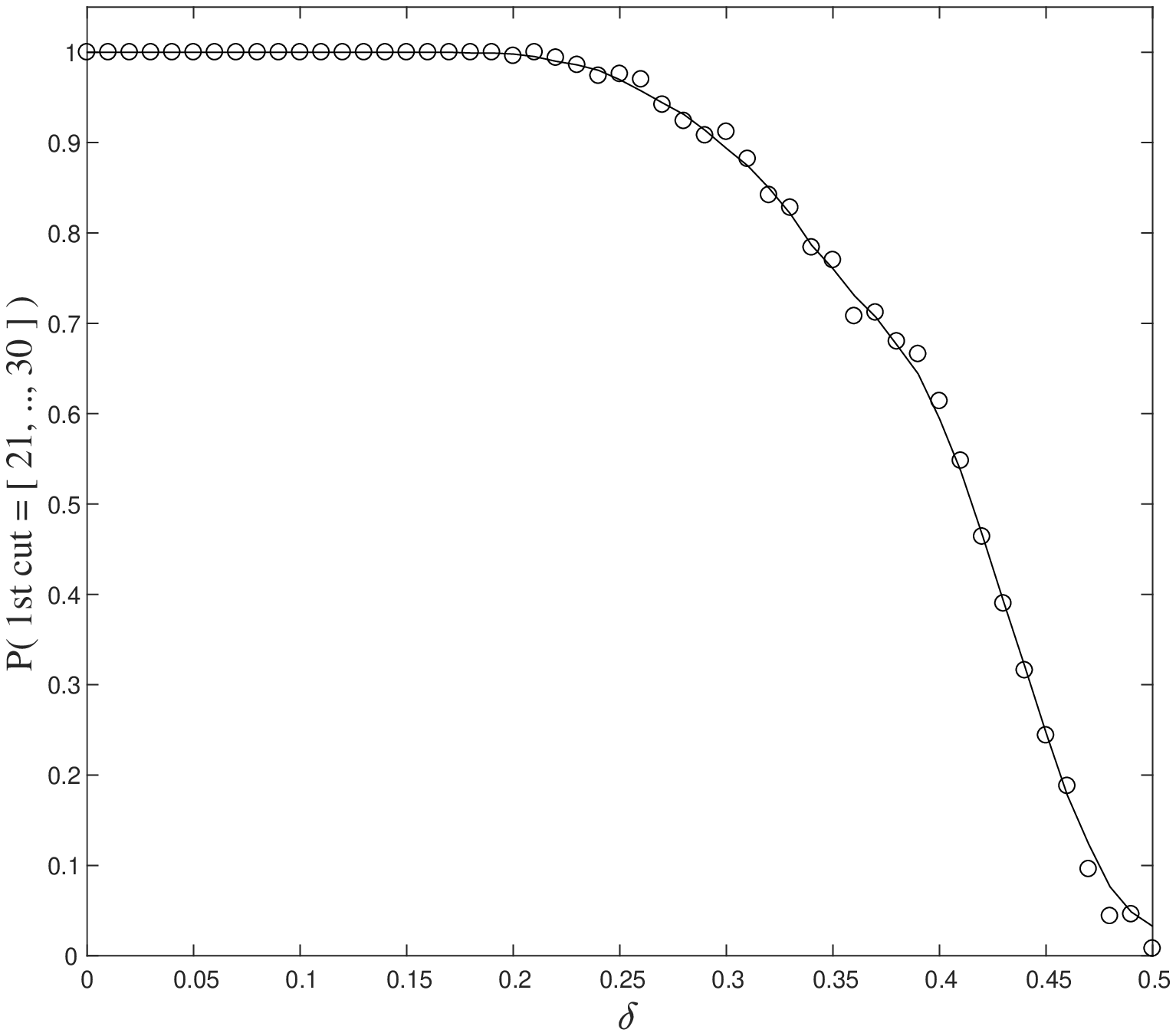}
\caption{\label{fig-nonlinear} The probability of the subset with the smaller number of elements in the  MIP is $\{21, 22, \ldots, 30\}$ is plotted as a function of the connection parameter $\delta \in [0, 1/2]$. Each circle shows the sample probability of 500 independent simulations, and the solid line shows the moving average of the probability.}
\end{figure}

To summarize, this case study confirmed our theoretical expectation that the MIP captures the block-wise informational components; namely that the partition probability is a decreasing function of the connection parameter $\delta$ (Figure \ref{fig-nonlinear}). This means that the MIP search detects the weakest underlying connection between 20 and 21, and successfully separates it into the two subsets, if the connections between 20 and 21 are weak. 

\section{Discussion \label{sec-discussion}}
In this paper, we proposed a fast and exact algorithm which finds the weakest link of a network at which the network can be partitioned with the minimal information loss (MIP). Since searching for the  MIP has the problem of combinatorial explosion, we employed Queyranne's algorithm for a submodular function. We first showed that the mutual information is a symmetric submodular function. Then, we used it as an information loss function to which Queyranne's algorithm can be applied. Our numerical case studies demonstrate the utility of the MIP search for a set of locally interacting nonlinear oscillators. This demonstration opens the general use of the MIP search for system neuroscience as well as other fields.

The proposed method can be utilized in Integrated Information Theory (IIT). \ADD{In IIT, information loss is quantified by integrated information. To date,} there are several variants of integrated information \cite{Tononi2004,Balduzzi2008,Oizumi2014,Barrett2011,Ay2015,Oizumi2016PLoS,Tegmark2016,Oizumi2016}. The mutual information was used as a measure of integrated information in the earliest version of IIT \cite{Tononi2004}, but different measures which take account of dynamical aspects of a network were proposed in the later versions \cite{Balduzzi2008,Oizumi2014}. To apply the proposed method, we first need to assess whether the other measures of integrated information are submodular or not. Even when the measures are not strictly submodular, the proposed algorithm may provide a good approximation of the MIP. An important future work is to assess the submodularity of the measures of integrated information, and also the goodness of the proposed algorithm as an approximation. 

\DEL{In the present study, we limit our focus to bi-partition to simplify the problem. There are some cases, however, for which tri-partition or general $k$-partition is considered appropriate. In general, the sequential bi-partitioning procedure does not necessarily result in the same partitioning for a given set of elements. Thus, a proper optimization procedure that is efficiently computable would in future be formulated in the form of recursive use of submodular optimization.}

\section*{Supporting information: Queyranne's algorithm \label{sec-Queyranne}}

In this appendix, we briefly describe Queyranne's algorithm \cite{Queyranne1998}. Suppose we have a submodular system $(V, g)$ where $V$ is a given set of elements and $g$ is a submodular function defined for the power set of $V$. Let $f( U ) := g( U ) + g( V \setminus U )$ for every subset $U \subseteq V$ be a symmetric function constructed with the submodular function $g$. Queyranne's algorithm is used to search the subset $U$ which minimizes the symmetric submodular function $f( U )$. \DEL{For example, in this study, we consider the case that $f(U)$ is mutual information (Eq. \ref{eq:MI}) and $g(U)$ is entropy.}
For example, in this study, we consider the case that $f(U) = 2I(U; V\setminus U)$, identified up to a constant multiplier, and $g(U) = I(U; V\setminus U)$ are both mutual information (Eq. \ref{eq:MI}).

In the algorithm proposed in \cite{Queyranne1998}, the key observation is that a special ordered pair $(t, u)$, called a {\em pendent pair}, can be identified for an arbitrary subset $U \subseteq V$ in $O(N^2)$ time. Identification of a pendent pair $(t, u)$ of the set $V$ reduces the search space because for the desired subset $U$ minimizing $f(U)$, either case (1) $U = \{u\}$ or (2) $U \supseteq \{ t, u \}$ holds. Thus, by keeping case (1) as a candidate for the minimal partition, we can further refine case (2), in which we define a new ground set $V'$ where the elements $\{t, u\}$ are treated as an inseparable unit element $u'$. By using the new merged element $u'$, $V'$ is defined as
\[
V' := \{V \setminus \{t, u\} \} \cup  \{ u' \}.
\]
After this procedure is applied once, the effective number of elements is reduced to $N-1$. By applying this procedure recursively to search the set $V'$ with $N-1$ elements, we would obtain another candidate for the minimal partition and a candidate set $V''$ with $N-2$ elements for further search. Thus, by finding the pendent pair for the given set $V$ at each step recursively, we obtain $N-1$ candidates for the minimal partition, and then find the minimal one from among them. In summary, this recursive computation takes $O(N^3)$ time because it requires the construction of a series of pendent pairs in $O(N^{2})$, and \Com{$N-1$ pendent pairs are needed to construct for minimization}.

Next we illustrate the construction of a pendant pair. An ordered pair $(t, u)$ of elements of $V$ is called a pendent pair for $(V, g)$, if $f( {u} )$ takes the minimum in all subsets of $V$ which separate $t$ from $u$, or equivalently
\[
 f( u ) = \min \{ f(U) \mid U \subset V, t \not\in U \text{ and } u \in U \}.
\]
There is at least one pendent pair for any symmetric submodular function. Further, a pendent pair can be constructed specifically for an element $x \in V$ as follows. For an element $x \in V$, let us write $v_{1} := x$, $W_{0} = \emptyset$, and $W_{1} = \{ v_{1} \}$.
For $i > 1$, 
\[
 v_{i} := \argmin_{u \in V \setminus W_{i-1}} g( W_{i-1} \cup \{u\}) - g( \{u\} ),
\]
and $W_{i} := W_{i-1} \cup \{ v_{i} \}$.
For a set $V$ of the size $N = |V|$, the $(v_{N-1}, v_{N})$ is a pendent pair.
This construction of a pendent pair needs $O( N^2 )$ times of evaluation of the function $f$.
Importantly, for all $y \in V \setminus W_{i}$ and all $x \subseteq W_{i-1}$ in the series $( W_{i} )_{i=1}^N$ constructed by the procedure above for the submodular system $(V, f)$, the following inequality holds
\[
 g( W_{i} ) + g( y ) \le g( W_{i} \setminus X ) + g( X + y ).
\]
See \cite{Queyranne1998} for the proof of this inequality. By putting $i=N-1$ in the inequality, we can see that the partition $(v_{N}, V \setminus \{v_{N}\})$ gives the minimum among all subsets separating $v_{N}$ from $v_{N-1}$.

By definition of the pendent pair, one of the following two cases, case 1 or 2, holds for a given pendent pair $(t, u)$.
\begin{enumerate}
 \item { The set $\{u\}$ is a solution of the minimization problem.}
 \item { Some set $U \supseteq \{ t, u \}$ is a solution of the minimization problem.}
\end{enumerate}
In the first case, the algorithm reports it.
In the second case, the algorithm constructs another submodular system $(V', f)$,
in which a new element is defined by merging the pendent pair $u' = \{t, u\}$
and $V' = V \setminus \{ t, u\} \cup u'$.
The new system $(V', f)$ with the merged pair is also submodular, and thus the same argument for the pendent pair can apply recursively.


\section*{Supporting information 2: Extension to $k$-partition algorithm}
The Queyranne's algorithm works on minimization of $g( U ) = f( U ) + f( V \setminus U)$ with respect to non-empty set $U \subset V$ or $g( (U, V\setminus U) = f(U) + f(V\setminus U )$ over bi-partition $(U, V\setminus U)$ with an arbitrary submodular set function $f$.
Here we show a recursive method extending this symmetric submodular search over a set of bi-partitions to that of $3$-partitions. The following argument will be easily extended to that of $k$-partition.

First let us denote the set of $k$-partitions of a given set $V$ by 
$$P_{k,V} := \left\{ (M_{0}, M_{1}, \dots, M_{k-1}) | \bigcup_{i}M_{i} = V, M_{i} \cap M_{j} = \emptyset \text{ for any } i \neq j\text{ and, } M_{i} \neq \emptyset \text{ for every } i \right\}.$$ 
For a submodular system $(V, f)$ of a given underlying set $V$ and a submodular set function $f: 2^{V} \mapsto \mathbb{R}$, 
we consider minimization of function $g: P_{3, V} \mapsto \mathbb{R}$ of the form
\begin{equation}
\label{eq-k-partition}
 g\left( ( M_{0}, M_{1}, M_{2} ) \right) = \sum_{i=0}^{2}f(M_{i}) + c,
\end{equation}
where $c \in \mathbb{R}$ is a constant. This is an extension of the bi-partition function $g( (U, V\setminus U) = f(U) + f(V\setminus U )$ to $3$-partition function. In this section, we provide an algorithm to minimize this $k$-partition function by employing Queyranne's algorithm.

By defining $f(M) := H(M)$ for $M \subseteq V$, $(V, f)$ is a submodular system, and 
the information loss function is written with a constant $c = -f(V)$ by 
$$
g\left( ( M_{0}, M_{1}, M_{2} ) \right) = \sum_{i=0}^{2}f(M_i) - f(V).
$$
For the special case $k=2$, $g( (M_{0}, M_{1} ) ) = I( M_{0}; M_{1} )$, this is identical to the minimal loss of information introduced in this study.

Our argument below does not depend on any specific form of a particular submodular function $f$, as long as the objective function takes the form in (\ref{eq-k-partition}).
The basic idea is to reduce the original objective function $g: P_{3, V} \mapsto \mathbb{R}$ to 
a set function $g_{3, V}: 2^{V} \mapsto \mathbb{R}$ by recursively defining $g_{2, U}$ for the remaining two subsets in a given bi-parition.
As our goal is to minimize $g_{3,V}$, such reduction can be written specifically for non-empty $U \subset V$ by 
\begin{equation}
\label{eq-recursive-def}
g_{3,V}( U ) := 
f_{V}( U ) + h_{2, V}( U ), 
\end{equation}
where 
for any $\emptyset \subset U_{1} \subset U_{2}$, 
\REP{$f_{U_{2}}( U_{1} ) := f(U_{1}) + f( U_{2} \setminus U_{1} )$}{$f_{U_{2}}( U_{1} ) := f(U_{1}) + f( U_{2} \setminus U_{1} ) - f( U_{2} )$}
and $h_{1, U_{2}}( U_{1} ) := 0$ and 
\begin{equation}
\label{eq-h}
h_{2, U_{2}}( U_{1} ) := 
\begin{cases}
  \min
    \begin{cases}
      \min_{ \emptyset \subset U' \subset U_{2} \setminus U_{1} }g_{2, U_{2} \setminus U_{1}}( U' )
      \\
      \min_{ \emptyset \subset U' \subset U_{1} }g_{ 2, U_{1} }( U' )
    \end{cases}
 & ( \min( |U_{1}|, |U_{2} \setminus U_{1}| ) > 1 )
\\
  \min_{ \emptyset \subset U' \subset U_{2} \setminus U_{1} }g_{2, U_{2} \setminus U_{1}}( U' ) & ( |U_{1}| = 1 )
\\
  \min_{ \emptyset \subset U' \subset U_{1} } g_{2, U}( U' ) & ( |U_{2} \setminus U_{1}| = 1 )
\end{cases}.
\end{equation}
This function (\ref{eq-recursive-def}) can be interpreted as recursive bi-partitioning across multiple stages: The first partition $(U, V \setminus U)$ of the set $V$ is made on $f_{V}$, and the second 
partition $(U', \overline{U}')$ of either $U$ or $V \setminus U$ on $h_{k-1}$, and so forth. For $|U|=1$ or $|V\setminus U|=1$, there is only one set for which the second partition can be made,
otherwise smaller one of either $f_{U}(M_{0})$ or $f_{V \setminus U}(M_{0})$ has the solution. 
For $k=2$, $g_{2, V}( U ) = f_{ V }( U )$, and 
minimization of $f_{V}(U)$ over the set of bi-partitions of $V$ can be computed by the Queyranne's algorithm.

If this function $g_{3, V}$ is symmetric and submodular, we can apply the Queyranne's algorithm to this function at every recursive step above.
Then, the minimum of $g_{3, V}( \hat{U} )$ is identical to $g( ( \hat{U}, M_{0}, M_{1} ) )$ with 
the $3$-partition is $(\hat{U}, \hat{M}_{0}, \hat{M}_{1})$ such that
\[
\hat{U} = \argmin_{\emptyset \subset U \subset V}g_{3, V}(U) \text{ and }
 ( \hat{M}_{0}, \hat{M}_{1} ) = \argmin_{(M_{0}, M_{1}) \in P_{2, V\setminus \hat{U}}}h_{2}( M_{0}, M_{1} )
\]
or $(V \setminus \hat{U}, \hat{M}_{0}, \hat{M}_{1})$ such that
\[
\hat{U} = \argmin_{\emptyset \subset U \subset V}g_{3, V}(U) \text{ and }
 ( \hat{M}_{0}, \hat{M}_{1} ) = \argmin_{(M_{0}, M_{1}) \in P_{2, \hat{U}}}h_{2}( M_{0}, M_{1} ).
\]
As $g_{k, V}$ is obviously symmetric by definition, our main question now is whether it is submodular.
The lemma following states that it is submodular. 
\begin{lem} \label{lem-recursive}
For a given submodular system $(V, f)$, the function $g_{3, V}: 2^{V} \mapsto \mathbb{R}$ is submodular\REP{.}{, if $f$ is monotone increasing.}
\end{lem}
\REP{The proof of Lemma \ref{lem-recursive} needs Lemma 2 which is stated after the proof.}{The proof of Lemma \ref{lem-recursive} is given in \cite{HidakaArxiv}.}

\section*{Acknowledgments}
We thank Ryota Kanai for his comments and discussions on earlier versions of this manuscript. This work was partially supported by CREST, Japan Science and Technology Agency, and by the JSPS KAKENHI Grant-in-Aid for Scientific Research on Innovative Areas JP 16H01609 and  for Scientific Research B (Generative Research Fields) JP 15KT0013.

\nolinenumbers

\bibliographystyle{plos2015}
\bibliography{TotalCorrReferences}

%
%
%

\end{document}